\documentclass[pre,twocolumn,superscriptaddress]{revtex4}
\usepackage{amsfonts}
\usepackage{amsmath}
\usepackage{amssymb}
\usepackage{graphicx}

\begin{document}

\title{Fractional-order correlation imaging with thermal light}
\author{De-Zhong Cao}
\thanks{Corresponding Author: dzcao@ytu.edu.cn}
\affiliation{Department of Physics, Yantai University, Yantai 264005, Shandong Province,
China}
\author{Qing-Chen Li, Xu-Cai Zhuang, Cheng Ren}
\affiliation{Department of Physics, Yantai University, Yantai 264005, Shandong Province,
China}
\author{Su-Heng Zhang}
\affiliation{College of Physics Science $\&$ Technology, Hebei University, Baoding
071002, China}
\author{Xin-Bing Song}
\thanks{Email Address: xin-bing.song@weizmann.ac.il}
\affiliation{Department of Physics of Complex Systems, Weizmann Institute of Science,
Rehovot 76100, Israel}

\begin{abstract}
In thermal light ghost imaging, the correlation orders were usually positive
integers in previous studies. In this paper, we examine the fractional-order
moments, whose correlation order are fractional numbers, between the bucket
and reference signals in the ghost imaging system. The crucial step in
theory is to determine the precise relation between the bucket signals and
reference signals. We deduce the joint probability density function between
the bucket and reference signals by regarding the reference signals as an
array of independent stochastic variables. In calculating the
fractional-order moments, the correlation order for the reference signals
must be positive to avoid infinity. While the correlation order for the
bucket signals can be positive or negative numbers. Negative (positive)
ghost images are obtained with negative (positive) orders of the bucket
signals. The visibility degree and signal-to-noise ratio of ghost images
from the fractional-order moments are analysed. The experimental results and
numerical simulations meet our analysis based on probability theory.
\end{abstract}

\maketitle

\section{Introduction}

Measuring high-order intensity correlation function is the key tool to
reconstruct the object information in ghost imaging with thermal light
(GITL) \cite{r0,r2,r3,r4,r5,r6,wtj1}. The correlation orders were natural
numbers in all present scenarios. The most favorite order was $2$, and the
second-order correlation functions in GITL were widely investigated both in
theory and in practice. Since the source plays a role of a conjugate mirror
\cite{wtj1}, ghost imaging with thermal light can be implemented without
lenses \cite{wtj2,wtj3}. The ghost images in computational GITL can be
formed with bucket signals measured by only one single pixel-detector \cite%
{cgi1}. Now GITL were applied in remote sensing \cite{remote}, lidar \cite%
{remote2}, imaging encryption \cite{encryp}, and biomedical imaging \cite%
{bio}. Multi-color GITL has been investigated to discriminate wavelength
information \cite{color1}, and to reconstruct RGB information of the color
object \cite{color2}. Higher-order correlation functions were used to
enhance the visibility degree \cite{3gi2} and improve contrast-to-noise
ratios \cite{3gi3,3gi4} of the ghost images. Third-order GITL were also
applied to construct two ghost images \cite{3gi1}. Recent investigations
showed that ghost images can be formed in first-order correlation
measurements with thermal light \cite{1gi1}.

Nevertheless, the orders of natural numbers are relative rough parameters in
application. Besides integer-order moments, the fractional-order moments
made great success in such processes as truncated L\'{e}vy flights \cite%
{levy} or atmospheric laser scintillations \cite{scill}. In this paper, we
report a GITL experiment in which the fractional-order moments of the
stochastic bucket and reference intensity signals are calculated. That is,
the object information is reconstructed by measuring the fractional-order
moments of the bucket and reference signals. In calculating the
fractional-order moments, the positive orders of the reference signals are
set to avoid infinity, while the orders of the bucket signals can be
positive or negative numbers. We find that negative (positive) ghost images
can be obtained with negative (positive) orders of the bucket signals.

In theory, an elaborate analysis based on probability theory is provided.
The crucial step is to determine the precise relation between the bucket
signals and reference signals. The reference signals can be regarded as an
array of independent stochastic variables, each of which meets negative
exponential distribution \cite{3gi3}. So the probability density function of
the bucket signals, as well as the joint probability function between the
bucket and reference signals, can be obtained since the bucket signals can
be regarded as a linear sum of the reference signals. Also the visibility
degrees and signal-to-noise ratios of the ghost images are analyzed
according to our theory. The experimental results and numerical simulations
are in good agreement with our theory.

Our paper is organised as follows. Section II gives the theory of the joint
probability density function between the bucket and reference signals.
Section III shows the theoretical analysis and experimental results of the
fractional-order moments for binary objects. Section IV shows the numerical
simulations of the fractional-order moments for a complicated object. The
conclusions and discussions are shown in Sec. V.

\section{Joint probability density function between the bucket and reference
signals}

Figure 1 shows the sketch of our experimental setup of GITL. The two sets of
correlated random speckles are the thermal light fields in the object plane
and reference plane. In the object arm, the bucket detector D$_{\text{B}}$
converts the total optical intensity out of the object, depicted by the
letter \textquotedblleft A\textquotedblright , into bucket signal $I_{B}$.
While the reference detector D$_{\text{r}}$ scans and converts local
intensity into reference signal $I_{r}$. The bucket detector and the
reference detector are two charge coupled devices (CCDs) in experiment. The
correlator, which is a computer indeed, is used to measure fractional-order
moment functions $\left\langle I_{B}^{\mu }I_{r}^{\nu }\right\rangle $,
where $\mu $\ and $\nu $ are fractional numbers. The fraction-order moment
function retrieves the object information which is shown in the screen.

\begin{figure}[tbh]
\centering \includegraphics[clip,width=7.5cm]{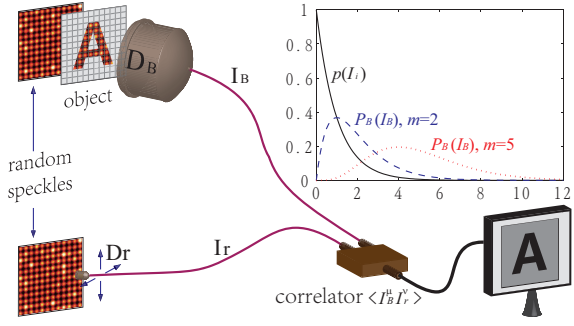}
\caption{Sketch of the experimental setup. The random speckles are two
identical pseudo-thermal light beams. The upper one is the object beam and
the lower is the reference beam. $D_{B}$ is the bucket detector,and the
pixel detector $D_{r}$ registers the reference beam. The fractional moments $%
\left\langle I_{B}^{\protect\mu }I_{r}^{\protect\nu }\right\rangle $ is
performed by the correlator.}
\end{figure}

The resolving power of GITL is inversely proportional to the coherence
length of the optical fields in the object and reference planes. Throughout
the paper we consider the case of perfect GITL that the GITL system has the
ability to completely reconstruct the object information. For simplicity of
mathematics, as shown in Fig. 1, we synchronically divide the object and
reference planes into such $n$ small units that (i) the details of the
object are maintained, and (ii) the thermal fields in all the units are
statistically independent from each other.

From the viewpoint of probability theory, the reference signals, i.e., the
thermal light intensities in all the units, can be regarded as a set of
stochastic variables $I_{r}=\{I_{1},I_{2},\cdots ,I_{n}\}$. Each element of
the reference signal $I_{i}$ ($i=1,2,\cdots ,n$) meets the negative
exponential probability distribution $p(I_{i})=\frac{1}{I_{0}}e^{-\frac{I_{i}%
}{I_{0}}}$, where the constant $I_{0}$ represents the intensity average. The
fractional moment of the reference signal is $\left\langle I_{i}^{\nu
}\right\rangle =\Gamma (1+\nu )I_{0}^{\nu }$ for any\ fractional number $\nu
$, where the Gamma function is $\Gamma (1+\nu )=\int_{0}^{\infty }t^{\nu
}\exp [-t]dt$. Due to the fact that the largest probability of the reference
signal is $p(0)$ that $p(0)\geqslant p(I_{i})$, we usually set $\nu $
positive ($\nu >0$) in experiment to avoid infinity.

The bucket detector D$_{\text{B}}$, which collects the variables scattered
from the object, output the bucket signals
\begin{equation}
I_{B}=I_{r}T=\sum\nolimits_{i=1}^{n}I_{i}t_{i},  \label{a1}
\end{equation}%
where $0\leq t_{i}\leq 1$ is the transmittance or reflectivity of the $i$th
unit in the object $T=\{t_{1},t_{2},\cdots ,t_{n}\}^{\prime }$ (the prime
denotes matrix transposition). Obviously, the bucket signal $I_{B}$ in Eq. (%
\ref{a1})\ is a linear sum of $n$ independent variables, each of which
fulfills the negative exponential probability density function with weight $%
t_{i}$. The probability density function for the variable from the $i$th
object unit becomes $\frac{p(I/t_{i})}{t_{i}}=\frac{1}{I_{0}t_{i}}e^{-\frac{%
I_{i}}{I_{0}t_{i}}}$, and its Laplace transformation is%
\begin{equation}
\widetilde{p}_{i}(s)=\int_{0}^{\infty }\frac{p(I/t_{i})}{t_{i}}e^{-sI}dI=%
\frac{1}{1+s\times I_{0}t_{i}}.  \label{a3}
\end{equation}
We can see that each variable from the object unit also fulfill the negative
exponential distribution, with a modified average $I_{0}t_{i}$. After some
algebra, the probability density function of the bucket signal is calculated
out%
\begin{equation}
P_{B}(I_{B})=\underset{s\rightarrow I_{B}}{\mathcal{L}^{-1}}\left[
\prod\nolimits_{i=1}^{n}\widetilde{p}_{i}(s)\right] ,  \label{a2}
\end{equation}%
where $\underset{s\rightarrow I_{B}}{\mathcal{L}}^{-1}$ denotes the inverse
Laplace transformation from variable $s$ to $I_{B}$.

We note that $P_{B}(I_{B})$ in Eq. (\ref{a2}) does not completely depend on
the object structure. The probability density function $P_{B}(I_{B})$ can be
calculated out as long as the histogram of the object is known. A remarkable
feature of the bucket probability density function is that $P_{B}(0)=0$, if
the object is composed of at least two nonzero units. So the fractional
moment of the bucket signal $\left\langle I_{B}^{\mu }\right\rangle
=\int_{0}^{\infty }I_{B}^{\mu }P_{B}(I_{B})dI_{B}$, where $\mu $ is a
fractional number, is tenable for both $\mu >0$ and $\mu <0$.

Since the reference signal $I_{i}$ is one constituent of the bucket signal $%
I_{B}$, we combine Eqs. (\ref{a1}) and (\ref{a2}), and write the joint
probability density function between the bucket signal $I_{B}$ and the
reference signal $I_{i}$ as
\begin{equation}
P_{2}(I_{B},I_{i})=P_{B}^{\prime }(I_{B}-I_{i}t_{i})\times p(I_{i}),
\label{a4}
\end{equation}%
where $P_{B}^{\prime }(x)=\underset{s\rightarrow x}{\mathcal{L}^{-1}}\left[
\prod\nolimits_{j=1,j\neq i}^{n}\widetilde{p}_{j}(s)\right] $, and $%
I_{B}\geq I_{i}t_{i}$.

The object information can be reconstructed by calculating out the
fractional-order moments $\left\langle I_{B}^{\mu }I_{r}^{\nu }\right\rangle
$ in a computer in experiment. According to the probability theory, the
function of the fractional-order moment is%
\begin{equation}
\left\langle I_{B}^{\mu }I_{i}^{\nu }\right\rangle =\int_{0}^{\infty
}dI_{B}\int_{0}^{I_{B}/t_{i}}dI_{i}\times I_{B}^{\mu }I_{i}^{\nu
}P_{2}(I_{B},I_{i}).  \label{a5}
\end{equation}%
As mentioned above, the fractional numbers in Eq. (\ref{a5}) meet $\nu >0$
and $\mu \neq 0$. What we should pay more attention to is the
fractional-order moments $\left\langle I_{B}^{\mu }I_{i}^{\nu }\right\rangle
_{0}$ and $\left\langle I_{B}^{\mu }I_{i}^{\nu }\right\rangle _{1}$ for $%
t_{i}=0$ and $1$, respectively. The former defines the background of the
ghost images $\left\langle I_{B}^{\mu }I_{i}^{\nu }\right\rangle
_{0}=\left\langle I_{B}^{\mu }\right\rangle \left\langle I_{i}^{\nu
}\right\rangle $. However, the latter meets%
\begin{equation}
\begin{array}{c}
\left\langle I_{B}^{\mu }I_{i}^{\nu }\right\rangle _{1}>\left\langle
I_{B}^{\mu }I_{i}^{\nu }\right\rangle _{0},\quad (\mu >0,\nu >0) \\
\left\langle I_{B}^{\mu }I_{i}^{\nu }\right\rangle _{1}<\left\langle
I_{B}^{\mu }I_{i}^{\nu }\right\rangle _{0}.\quad (\mu <0,\nu >0)%
\end{array}
\label{a6}
\end{equation}%
It is clear that the ghost image is above its background when $\mu >0$, and
is below its background when $\mu <0$. That is, negative ghost images can be
obtained for negative fractional order $\mu $. Consequently, the visibility
degree and peak SNR of the ghost images in fractional-order moments are
defined as%
\begin{equation}
\begin{array}{c}
V=\frac{\left\vert \left\langle I_{B}^{\mu }I_{i}^{\nu }\right\rangle
_{1}-\left\langle I_{B}^{\mu }I_{i}^{\nu }\right\rangle _{0}\right\vert }{%
\left\langle I_{B}^{\mu }I_{i}^{\nu }\right\rangle _{1}+\left\langle
I_{B}^{\mu }I_{i}^{\nu }\right\rangle _{0}}, \\
R_{p}=\frac{\sqrt{N}\left\vert \left\langle I_{B}^{\mu }I_{i}^{\nu
}\right\rangle _{1}-\left\langle I_{B}^{\mu }I_{i}^{\nu }\right\rangle
_{0}\right\vert }{\sqrt{\left\vert \left\langle I_{B}^{2\mu }I_{i}^{2\nu
}\right\rangle _{1}-\left\langle I_{B}^{\mu }I_{i}^{\nu }\right\rangle
_{1}^{2}\right\vert }},%
\end{array}
\label{a8}
\end{equation}%
where $N$ is the number of sampling in experiment \cite{3gi3}.

In the following, we show the experimental results of the fractional-order
moments for binary objects in ghost imaging, and then show the numerical
simulations of the fractional moments for a complicated object in ghost
imaging.

\section{Experiment results of binary objects}

To explicitly exhibit the probability theory method and to illustrate the
characteristics of ghost images from fractional-order moments in GITL, we
consider the case of binary objects that the values of the object units are $%
t_{i}=0$ or $1$. From Eq. (\ref{a2}), the bucket signals meet Gamma
distribution with probability density function as
\begin{equation}
P_{B}(I_{B})=\frac{I_{B}^{m-1}}{(m-1)!I_{0}^{n}}\exp [-\frac{I_{B}}{I_{0}}],
\label{b1}
\end{equation}%
where $m$ is the number of the effective object units (whose values are $%
t_{i}=1$). The subplot in Fig. 1 shows probability density functions of Eq. (%
\ref{b1}) for $p(I_{r})$ $m=1$ with solid black line, $P_{B}(I_{B})$ $m=2$
with dashed blue line, and $P_{B}(I_{B})$ $m=5$ with dotted red line,
respectively. We can find that $p(0)\geq p(I_{r})$ for single-unit object
and $P_{B}(0)\leq P_{B}(I_{B})$ for complex object. The joint probability
density function between the reference and bucket signals is%
\begin{equation}
P_{2}(I_{B},I_{i})=\left\{
\begin{array}{c}
\frac{(I_{B}-I_{i})^{m-2}}{(m-2)!I_{0}^{m}}\exp [-\frac{I_{B}}{I_{0}}%
],(t_{i}=1) \\
\frac{I_{B}^{m-1}}{(m-1)!I_{0}^{m+1}}\exp [-\frac{I_{B}+I_{i}}{I_{0}}%
],(t_{i}=0)%
\end{array}%
\right.  \label{b2}
\end{equation}%
where $I_{i}\leq I_{B}$ must be considered.

The ensemble average of the reference signals is $\left\langle
I_{i}\right\rangle =I_{0}$, and the ensemble average of the bucket signals
is $\left\langle I_{B}\right\rangle =mI_{0}$. The fractional-order moments
is calculated out%
\begin{equation}
\left\langle I_{B}^{\mu }I_{i}^{\nu }\right\rangle _{0}=\frac{\Gamma (m+\mu
)\Gamma (1+\nu )}{\Gamma (m)}I_{0}^{\mu +\nu },  \label{b3}
\end{equation}%
\begin{equation}
\left\langle I_{B}^{\mu }I_{i}^{\nu }\right\rangle _{1}=\frac{\Gamma (m+\mu
+\nu )\Gamma (1+\nu )}{\Gamma (m+\nu )}I_{0}^{\mu +\nu },  \label{b4}
\end{equation}%
for $t_{i}=0$ and $t_{i}=1$, respectively. The visibility degree and peak
SNR of the ghost images are%
\begin{equation}
V=\left\vert \frac{\Gamma (m+\mu +\nu )\Gamma (m)-\Gamma (m+\mu )\Gamma
(m+\nu )}{\Gamma (m+\mu +\nu )\Gamma (m)+\Gamma (m+\mu )\Gamma (m+\nu )}%
\right\vert ,  \label{b5}
\end{equation}%
\begin{equation}
R_{p}=\frac{\left\vert \frac{\Gamma (m+\mu +\nu )\Gamma (1+\nu )}{\Gamma
(m+\nu )}-\frac{\Gamma (m+\mu )\Gamma (1+\nu )}{\Gamma (m)}\right\vert }{%
\sqrt{\frac{\Gamma (m+2\mu +2\nu )\Gamma (1+2\nu )}{N\cdot \Gamma (m+2\nu )}-%
\frac{\Gamma ^{2}(m+\mu +\nu )\Gamma ^{2}(1+\nu )}{N\cdot \Gamma ^{2}(m+\nu )%
}}},  \label{b6}
\end{equation}%
respectively for binary objects.

In experiment, the pseudo-thermal light source is obtained by projecting a
laser beam (laser diode: $\lambda =650$nm) onto a rotating ground glass
plate \cite{3gi2} (which is not shown in Fig. 1). We set the diameter of the
laser spot on the glass plate $d=4.50$mm, and the distance between the
object and the glass plate $L=8.50cm$. The coherent length of the random
laser speckles in the object plane is about $\frac{1.22\lambda L}{d}\simeq
14.98\mu m$, which is smaller than the pixel pitch of the CCD $20.0\mu m$.
This ensures well-performed GITL, and fulfills the two assumptions proposed
above.

\begin{figure}[tbh]
\centering\includegraphics[clip,width=3.5cm,angle=-90]{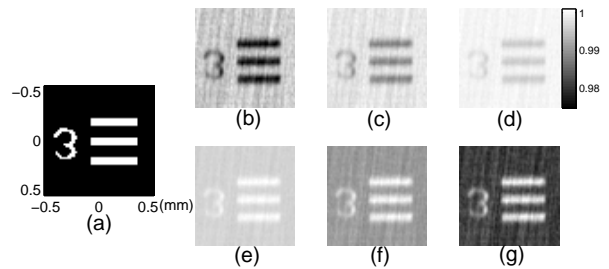}
\caption{Experimental results of ghost images from fractional-order moments
with binary objects. The fractional orders are set $\protect\mu =-2.7183$ in
(b), $\protect\mu =-1.414$ in (c), $\protect\mu =-0.618$ in (d), $\protect%
\mu =0.618$ in (e), $\protect\mu =1.414$ in (f), and $\protect\mu =2.7183$
in (g). The parameter $\protect\nu =0.5$ is fixed in all the ghost images.
The binary object is shown in (a).}
\end{figure}

Figure 2 shows our experimental results of measuring the fractional-order
moments in GITL with binary objects over $N=120,000$ samplings. The object
is depicted in Fig. 3(a). Figures 3(b-f) show the normalized
fractional-order moments $\frac{\left\langle I_{B}^{\mu }I_{i}^{\nu
}\right\rangle }{\left\langle I_{B}^{\mu }\right\rangle \left\langle
I_{i}^{\nu }\right\rangle }$. The ghost images from the fractional-order
moments are depicted in Figs. 2(b), 2(c), 2(d), 2(e), 2(f), and 2(g), for $%
\mu =-2.7183$, $-1.414$, $-0.618$, $0.618$, $1.414$, and $2.7183$
respectively. The parameter $\nu =0.5$ is fixed in all the ghost images. We
can see that the negative ghost images are obtained for negative orders $\mu
<0$ in Figs. 2(b), 2(c), and 2(d). While the positive ghost images are
obtained for positive orders $\mu >0$ in Figs. 2(e), 2(f), and 2(g). We find
that the greater the absolute order $\left\vert \mu \right\vert $ is, the
higher visibility degree of the ghost image becomes. In general, the
negative ghost images has better visibility than the positive ones. But the
behaviors of the signal-to-noise ratio (SNR) differs greatly from that of
the visibility degree. The SNRs decrease when $\left\vert \mu \right\vert $
increases. The Visibility degrees are $V=6.77\times 10^{-3}$, $3.45\times
10^{-3}$, $1.49\times 10^{-3}$, $1.47\times 10^{-3}$, $3.32\times 10^{-3}$,\
$6.26\times 10^{-3}$, and the peak SNRs (defined in Eq. (\ref{a8})) are $%
R_{p}=2.118$, $2.126$, $2.130$, $2.132$, $2.133$, $2.131$ for Figs. 2(b),
2(c), 2(d), 2(e), 2(f), and 2(g), respectively.

The orders $\mu $ and $\nu $\ determine the quality of ghost images.
Therefore, fractional orders provide more detailed parameters to exquisitely
adjust or modify the quality of the ghost images in practice than integer
orders. Below is our theory of the fractional-order moments in GITL.

The visibility $V$ in Eq. (\ref{b5}) and the relative peak SNR $R_{p}/\sqrt{N%
}$ in Eq. (\ref{b6}) versus the fractional orders are plotted in Fig. 3. The
number of effective object unit is $m=20$ in Figs. 3(a) and 3(b), and $m=30$
in Figs. 3(c) and 3(d), respectively. We can see from figs. 3(a) and 3(c)
that the visibility degree increases as the absolute values of the
correlation orders $\mu $\ and $\nu $ increase. An evident feature is that
the visibility degree of the negative image ($\mu <0$) increase faster than
that of the positive image ($\mu >0$). Also, the greater number of effective
object unit can degrade the visibility degree.

The relative peak SNR is plotted in Figs. 3(b) and 3(d). The peak SNR first
increases and then decreases as the correlation orders $\left\vert \mu
\right\vert $\ and $\nu $\ increase. We can also find that the maximum value
of peak SNR of the negative image\ ($\mu <0$) is greater than that of the
positive image ($\mu >0$). We can conclude that the image quality of the
negative ghost images can be better than that of the positive ghost images
for the opposite fractional orders. Furthermore, the visibility degree and
SNR of the ghost image vary continuously with fractional (continuous)
orders. So we can adjust the visibility degree and SNR at will by choosing
appropriate experimental parameters.

\begin{figure}[htb]
\centering\includegraphics[clip,width=5.5cm,angle=-90]{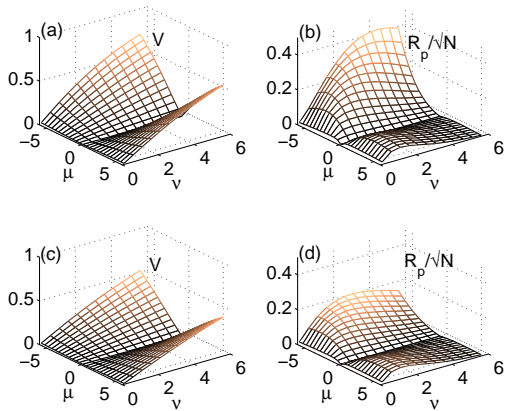}
\caption{Visibility degrees $V$ (a,c) and relative peak SNR $R_{p}/\protect%
\sqrt{N}$ (b,d) of binary objects. The object unit number (for $t_{i}=1$) is
$m=20$ in (a,b), and $m=30$ in (c,d).}
\end{figure}

\section{Numerical simulations of complex objects}

So far we have presented the theory of the fractional ghost imaging with
thermal light, and have illustrated ghost images in fractional-order moments
in experiment with binary objects. Our proposal of course is suitable for
more complicated objects. In this section we show the numerical simulations
of the fractional-order moments in GITL with a more complicated object.

The object is an image of the part of cameraman (size: $64\times 64$
pixels). Figure 4 shows the results of numerical simulations by calculating
the normallized fractional-order moments $\frac{\left\langle I_{B}^{\mu
}I_{i}^{\nu }\right\rangle }{\left\langle I_{B}^{\mu }\right\rangle
\left\langle I_{i}^{\nu }\right\rangle }$. The reference signals are the
stochastic numbers crated by a computer. The number of sampling is $%
N=200,000 $. Other parameters are the same as in Fig. 2. We again obtain
negative ghost images for $\mu <0$ in Figs. 4(a), 4(b) and 4(c). The
positive ghost images for $\mu >0$ 4(d), 4(e), and 4(f), are also obtained.
The visibility degrees are $V=7.84\times 10^{-4}$, $3.97\times 10^{-4}$, $%
1.76\times 10^{-4} $, $1.76\times 10^{-4}$, $4.06\times 10^{-4}$, $%
7.72\times 10^{-4}$, for Figs. 4(a), 4(b), 4(c), 4(d), 4(e), and 4(f),
respectively. Again the visibility increases when the absolute value of the
order increases. The corresponding peak SNRs for all the ghost images are $%
R_{p}=5.27$, $5.57$, $5.44$, $6.95$, $5.65$, and $6.18$, respectively. The
peak SNRs of the ghost images vary with the fraction orders.

\begin{figure}[tbh]
\centering \includegraphics[clip,width=5.0cm,angle=-90]{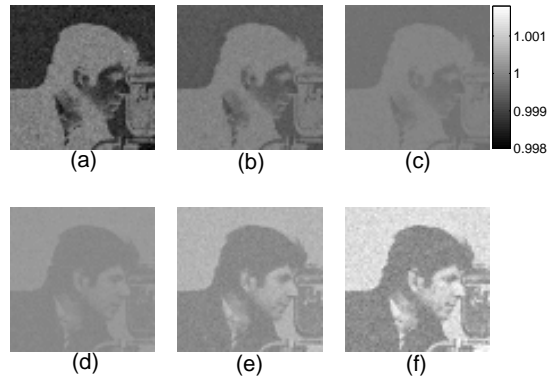}
\caption{Numerical simulations of ghost images from fractional moments with
a gray object ($64\times 64$ pixels). The number of sampling is $N=200,000$.
Other parameters are the same in Fig. 2.}
\end{figure}

\section{Conclusion}

In conclusion, we have investigated the fractional-order moments in GITL
experiment. The reference signals have been regarded as an array of
stochastic variables, while the bucket signals have been regarded as a
linear sum of these stochastic variables. We then have deduced the joint
probability density function between the bucket signals and reference
signals according probability theory. The object information can be
retrieved through fractional-order moments. We have found that negative
(positive) ghost images can be obtained if the orders of the bucket signals
are smaller (greater) than zero. Ghost imaging with fractional-order moments
has been implemented in experiments with a binary object and in numerical
simulations with a more complicated object. The visibility degrees and peak
SNRs of the ghost images vary with the correlation fractional-orders. So we
have the chance to carefully adjust the image quality by choosing
appropriate fractional orders in GITL. Our technique can provides abundant
ghost images, and has potential to work in complex environments.

This work was benefited from financial support by the National Natural
Science Foundation of China under Grant Nos. 11674273, 11304016, and
11204062.

\end{document}